\documentclass[twocolumn,showpacs,preprintnumbers,amsmath,amssymb]{revtex4}
\usepackage{graphicx}
\usepackage{amsmath}% Include figure files
\usepackage{dcolumn}% Align table columns on decimal point
\usepackage{hyperref}

\begin{document}

\title{Spatial light modulation based on coherent population oscillation in semiconductor quantum dots}% Force line breaks with

\author{Wen-Ming Ju}
\email{williamwallace @ sjtu.edu.cn}
\author{Ka-Di Zhu}%
\author{Huan Wang}%
\author{Pei-Hao Huang}%
\affiliation{%
Department of Physics, Shanghai Jiao Tong University, Shanghai
200240, P.R.China
}%

\date{\today}% It is always \today, today,

\begin{abstract}

Due to coherent population oscillation (CPO) effect, we
theoretically examine the generation and manipulation of
Laguerre-Gaussian beams in terms of phase and amplitude modulation
respectively in semiconductor quantum dots (SQDs). The results
indicate that both phase modulation with low absorption of probe
field and amplitude modulation with high efficiency can be achieved.
Thus the practical way demonstrates that the SQDs system can
function as effective optically addressed spatial light modulator
with which the transverse spatial properties of probe fields can be
artificially modulated.

\end{abstract}

\pacs{42.50.Gy; 78.67.Hc, 73.21.La, 03.67.-a}
\maketitle
 \vskip 2pc
\textbf{I INTRODUCTION}
 \vskip 1pc
A spatial light modulator (SLM) is an object that imposes some form
of spatially-varying modulation on a beam of light. It can modulate
the intensity as well as the phase of the light beam. Nowadays a
great deal of research~\cite{Pugatch,Shuker,Vudyasetu,Zhao} is
focused on storing information imprinted in the transverse plane of
the probe beam (i.e., images) and on reducing the effects of atomic
diffusion. Shuker \emph{et al.} pointed out that the immunity to
diffusion can be improved by appropriately manipulating the phase of
different points in the image~\cite{Shuker}. In the process, the
utilization of a SLM is necessary. In fact, SLMs are used
extensively in holographic data storage and display setups. They can
encode information into a laser beam and produce light fields with
peculiar spatial structures.

On the other hand, electromagnetically induced transparency (EIT)
has been intensively investigated for
decades~\cite{Fleischhauer,Chang-Hasnain}. Nevertheless, coherent
population oscillation (CPO), which is also a phenomenon of quantum
interference as EIT, has been proven to be another powerful
technique that can eliminate the absorption at the pump-probe beat
frequency and dramatically change the refractive
index~\cite{Zhao1,Ohman,Su}. In systems based on EIT and CPO, the
probe field can be manipulated coherently and all-optically by
strong pump field. With the well-known properties, these types of
systems are good candidates to realize velocity control of light,
storage and retrieval of images and other information processing.
Moreover, the slow light based on CPO can easily be achieved in a
solid-state material at room temperatures, while for EIT, it is
usually required to cool the medium to very low temperatures to
obtain a large index slope~\cite{Ohman}. Thus the CPO effect leads
itself more readily towards realistic applications using solid-state
devices, such as semiconductors.

In view of the advantages of CPO effect, in this paper, according to
CPO, we propose a way to spatially modulate probe light beams with
the use of optical patterns (e.g., images) with desired intensity
distributions in the pump fields. The material we introduce to
perform CPO consists of semiconductor quantum dots (SQDs) in which
excitons behave as two-level systems. To exemplify our proposal, we
examine the generation and manipulation of Laguerre-Gaussian (LG)
beams. The theoretical results suggest the feasibility of the
modulation of spatial light in terms of CPO in SQDs and our work may
open an avenue to SLM technology in the future.
 \vskip 2pc
\textbf{II THEORY}
 \vskip 1pc
Our model consists of tens of layers of SQDs embedded in a GaAs
matrix. We consider a spherical SQD in the presence of a strong pump
field and a weak probe field. For the SQD, we assume a two-level
model which consists of the ground state $|0\rangle$ and the first
excited state (single exciton) $|1\rangle$. The SQD via exciton
interacts with a probe field of frequency $\omega_s$ and is
coherently driven by a strong pump field of frequency $\omega_c$. As
usual, this two-level system can be characterized by the
pseudo-spin-$\frac{1}{2}$ operators $S^\pm$ and $S^z$. Then the
Hamiltonian of the system in a rotating frame at the pump field
frequency $\omega_c$ reads as follows:

\begin{equation}
\begin{split}
H=&\hbar(\omega_{ex}-
\omega_c)S^z-\hbar\Omega(S^++S^-)/2\\&-\mu(S^+E_s e^{-i\delta
t}+S^-E^*_s e^{i\delta t})/2.
\end{split}
\end{equation}
where $\hbar \omega_{ex}$ is the energy of exciton binding,
$\Omega=\mu E_c/\hbar$ is the Rabi frequency, where $\mu$ is the
interband dipole matrix element, $E_c$ is the slowly varying
envelope of the pump field. $E_s$ is the slowly varying envelope of
the probe field. $\delta=\omega_s-\omega_c$ is the detuning of the
probe and the pump field.

The temporal evolution of the exciton in the SQD are determined by
the Heisenberg equation of motion. After replacing the operator by
the mean values defined as classical variable $\langle S^-\rangle$
and $\langle S^z\rangle$ and then setting $w=2\langle S^z\rangle$
and $p=\mu\langle S^-\rangle$~\cite{Agarwal,Greene,Lam}, we have the
generalized optical Bloch equations:

\begin{equation}
\frac{dp}{dt}=-(\frac{1}{T_2}+i\Delta)p-\frac{i\mu^2}{2\hbar}wE,
\end{equation}

\begin{equation}
\frac{dw}{dt}=-\frac{1}{T_1}(w+1)+2\texttt{Im}(pE^*)/\hbar.
\end{equation}
where $\Delta=\omega_{ex}-\omega_c$ and $E=E_c+E_s e^{-i\delta t}$.
$T_1$ is the exciton lifetime and $T_2$ is the exciton dephasing
time.

On making the ansatz~\cite{Boyd}: $p=p_0+p_1 e^{-i\delta
t}+p_{-1}e^{i\delta t}, w=w_0+w_1 e^{-i\delta t}+w_{-1}e^{i\delta
t}$, obtaining and solving the equations in the steady state, the
dimensionless susceptibility is given by

\begin{equation}
\begin{split}
\chi^{(1)}(\omega_s)=&\frac{p_1}{E_s\mu^2
T_2}=\frac{-iw_0}{1+i(\Delta_c-\delta_c)}\times\\&\{1-\frac{2\Omega_c^2}{D(\delta_c)}(1+i\Delta_c)
[1-i(\Delta_c+\delta_c)](2-i\delta_c)\},
\end{split}
\end{equation}
where
$D(\delta_c)=(\frac{T_2}{T_1}-i\delta_c)(1+\Delta_c^2)\times[(1-i\delta_c)^2+\Delta_c^2]+4\Omega_c^2(1-i\delta_c)(1+\Delta_c^2)$,
$\delta_c=\delta T_2$, $\Omega_c=\mu\frac{E_c}{\hbar}T_2$,
$\Delta_c=\Delta T_2$, and the population inversion of exciton $w_0$
is determined by
$w_0=\frac{-4\Omega_c^2w_0\frac{T_1}{T_2}}{1+\Delta_c^2}-1$. Also,
we can have $\Delta_s=(\omega_s-\omega_{ex})T_2$ which is the
frequency detuning between the probe field and exciton. Note that
the Rabi frequency $\Omega$ and the detunings are all normalized
with respect to the exciton dephasing time $T_2$.
 \vskip 1pc
\begin{figure}[h]
\includegraphics[width=0.4\textwidth]{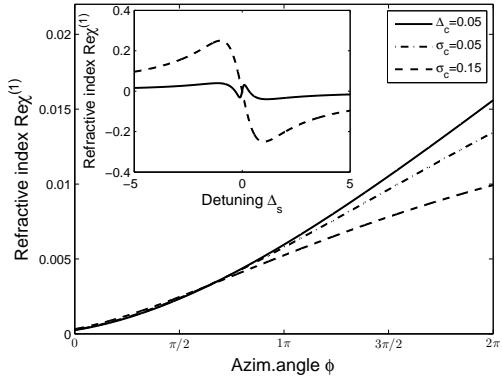}
\caption{The real part of the linear susceptibility for the probe
field induced by the azimuthal intensity distribution for three
cases: a fixed detuning $\Delta_c=0.05$ (solid line); Gaussian
distribution: $\sigma_c=0.05$ (dash-dotted line), $\sigma_c=0.15$
(dashed line). The frequency detuning between the probe field and
exciton is $\Delta_s=0$. The inset shows the real part of the linear
susceptibility for the probe field as a function of $\Delta_s$ with
dashed line showing Re$\chi^{(1)}$ for zero pump field, with solid
line showing Re$\chi^{(1)}$ for the Rabi frequency of the pump field
$\Omega_c=0.3$. $\Delta_c$ is fixed as 0.05.} \label{fig1}
\end{figure}

\begin{figure}[t]
\includegraphics[width=0.4\textwidth]{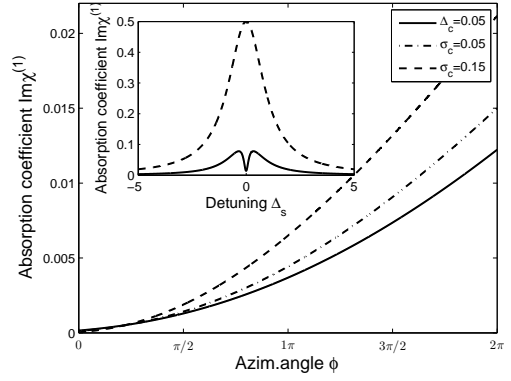}
\caption{The imaginary part of the linear susceptibility for the
probe field induced by the azimuthal intensity distribution for
three cases: a fixed detuning $\Delta_c=0.05$ (solid line); Gaussian
distribution: $\sigma_c=0.05$ (dash-dotted line), $\sigma_c=0.15$
(dashed line). The detuning $\Delta_s=0$. The inset shows the
imaginary part of the linear susceptibility for the probe field as a
function of $\Delta_s$ with dashed line showing Im$\chi^{(1)}$ for
zero pump field, with solid line showing Im$\chi^{(1)}$ for the Rabi
frequency of the pump field $\Omega_c=0.3$. $\Delta_c$ is fixed as
0.05.} \label{fig2}
\end{figure}

 \vskip 2pc
\textbf{III RESULTS AND DISCUSSION}
 \vskip 1pc
Since the dimensionless linear optical susceptibility of our system
has been derived, the refractive index (associated with
Re$\chi^{(1)}$) and the absorption spectrum (associated with
Im$\chi^{(1)}$) as functions of the detuning $\Delta_s$ can be
obtained and are shown in the insets of Fig.~\ref{fig1} and
Fig.~\ref{fig2}, respectively.  Here and in the following
calculation, for the relaxation time in the SQD in the
room-temperature regime, we take $T_2=3\times10^{-13}$ s,
$T_1=1.5\times10^{-11}$ s~\cite{Jiang}, and choose $\Omega_c=0.3$,
$\Delta_c=0.05$. It can be seen clearly that when the pump field is
turned on, the slope of the refractive index is dramatically changed
at the resonant condition $\Delta_s=0$. In addition, a non-absorbing
hole appears at $\Delta_s=0$, suggesting the system becomes
transparent for the probe field in the presence of pump field. Now
let's consider the generation of LG modes by means of phase
modulation. Firstly we give a chief introduction of the possible
experimental setup. The probe beam is shaped as a Gaussian beam and
is in exact resonance with the transition
$|0\rangle\longleftrightarrow|1\rangle$ (i.e., $\Delta_s=0$). The
pump beam passes through a image mask. The plane of the image mask
is imaged onto the center of the GaAs matrix by using lens. The pump
and probe beams are combined on a beam splitter and copropagate
through the SQDs. After that, they are divided by a second beam
splitter. To generate LG modes, an azimuthal phase winding
$e^{il\phi}$ should be imprinted onto the wave front of the incident
Gaussian probe field, where $l\phi$ is the helical phase, $l$ is the
integer winding number and $\phi$ is the azimuthal angle around
optical axis~\cite{Grier}. The phase imprinting leads to an
azimuthal variation of the refractive index $n(\phi)$ in our system.
To realize it, an azimuthally dependent Rabi frequency
$\Omega_c(\phi)$ of the pump field is required which can make
$\chi^{(1)}$ vary with $\phi$ according to Eq.(4). \begin{figure}[t]
\includegraphics[width=0.45\textwidth]{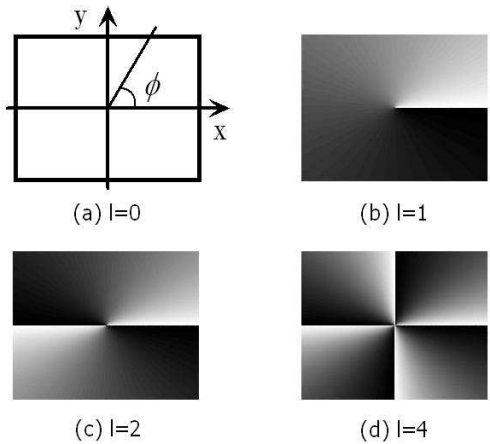}
\caption{(a) Uniform illumination of the pump field with the integer
winding number $l=0$; (b) Azimuthal intensity distribution of the
pump field to generate the LG beam with $l=1$, the intensity
distribution is $I(\phi)\propto\frac{a}{b\phi+c}$; (c) $l=2$; (d)
$l=4$.} \label{fig3}
\end{figure} Thus we
assume~\cite{Zhao2}

\begin{equation}
\Omega_c(\phi)=\frac{\mu E_c}{\hbar}T_2=\sqrt{\frac{a}{b\phi+c}},
\end{equation}
where $a$, $b$, $c$ are the adjustable parameters. This pump field
can be constructed by the images of amplitude masks. To illustrate
clearly, the intensity distribution $I(\phi) (\propto E_c^2(\phi))$
of the pump field is shown in terms of images in case of $l=0,1,2,4$
respectively in Fig.~\ref{fig3}. One can see uniform illumination
when $l=0$ because of disappearance of phase shift. One can also
find the fold of intensity distribution with higher winding number
as compared to the image with $l=1$. In what follows, we assume
$a=1$, $b=c=1$ to make the pump field strong:
$\Omega_c(\phi)\gg1/T_1$, take the detuning $\Delta_c=0.05$. The
solid lines in Fig.~\ref{fig1} and Fig.~\ref{fig2} show the real
part and imaginary part of $\chi^{(1)}$ for the probe field in the
SQDs as functions of azimuthal angle $\phi$, respectively. We can
see that the absorption of the system is suppressed evidently within
a $2\pi$ phase regime. In the theoretical calculations above, the
detuning $\Delta_c$ is a fixed value, which can only stand for the
condition of one quantum dot with a certain dot size. To describe
the ensemble of SQDs and reflect the size distribution of the dots
and the inhomogeneous broadening of the transitions, we assume a
Gaussian distribution function for the pump field-exciton detuning
$\Delta_g$:

\begin{equation}
G(\Delta_g)=\frac{1}{\sqrt{2\pi}\sigma_c}\exp[-\frac{(\Delta_g-\Delta_c)^2}{2\sigma_c^2}],
\end{equation}
where $\Delta_c$ is the center of Gaussian wave-packets and
$\sigma_c=\sigma T_2$ is the half-width of the distribution. Thus,
more precise results can be obtained for Re$\chi^{(1)}$ and
Im$\chi^{(1)}$ and they are plotted with dash-dotted and dashed
lines in Fig.~\ref{fig1} and Fig.~\ref{fig2} to present a comparison
with the former results.

Furthermore, since a $2\pi$ phase difference between the maximum and
minimum should be imprinted onto the wave front of the probe field,
we can obtain~\cite{Zhao2}: $\Delta n\cdot d=[n(2\pi)-n(0)]d=\Delta
l\cdot\lambda$, where $n$ is the refractive index, $d$ is the
thickness of the SQDs medium, and $\Delta l=1$ because we discuss,
for simplicity, the generation of LG mode with $l=1$ from the
Gaussian probe field with $l=0$. In the CPO system, it is easy to
obtain $n\approx1+\frac{1}{2}\chi^{'}$, which can result in
$d=\frac{2\lambda}{\texttt{Re}\chi^{(1)}(2\pi)-\texttt{Re}\chi^{(1)}(0)}$,
where $\lambda\approx\frac{2\pi c\hbar}{E}=530$ nm, and $E=2.34$ eV
is the interband transition energy~\cite{Jiang}. Thus we obtain the
thickness of the solid medium $d\approx71$ $\mu$m. With the
thickness $d$, we can calculate the transmission probability
$T(=e^{-\alpha d})$, where the absorption coefficient
$\alpha=2\pi$Im$\chi^{(1)}/\lambda$. In most cases, the transmission
probability of the probe field in our system is above 80\% within a
$2\pi$ phase regime.

Secondly comes the discussion on amplitude modulation. Beams with
phase singularities such as LG beams have been generated at optical
wavelengths by the diffraction of a plane wave using specially
synthesized holograms or masks. Among the holograms or masks, the
forked binary grating is one kind of practical
application~\cite{Brand,Brand1}. The boundaries between the
transparent and opaque regions are given by a special instance of
the general formula, which can be expressed in polar coordinates
($r,\phi$) as~\cite{Brand1}:

\begin{equation}
p\frac{\phi}{\pi}=n+\frac{2r}{D}\cos\phi,
\end{equation}
where $p$ is the topological charge, $D$ is the period of the
grating far away from the forked center and $n=0,\pm1,\pm2,...$
Thus, to generate LG modes by means of amplitude modulation, the
aforementioned images with azimuthal intensity distributions should
be replaced by images with forked binary patterns~\cite{Zhao2}. As a
result, in the dark fringes of the forked pump field (i.e.,
$\Omega_c=0$), with the same detunings as those in phase modulation
(i.e., $\Delta_c=0.05, \Delta_s=0$), we can obtain Im$\chi^{(1)}=0.4
\Rightarrow T=e^{-\alpha d}=e^{-3.3\times 10^2}\rightarrow0$, where
$d=71$ $\mu$m is obtained through the previous calculation.
Nevertheless, in the bright fringes of the forked pump field where
the strong pump field is introduced, the outcome is $T\rightarrow1$,
which demonstrates the probe field can transmit with low absorption
in the system. Thus a tunable forked binary amplitude grating is
formed in the SQDs system for the probe field. The transmission
grating is made by first photographing a computer-generated pattern
and has been successfully demonstrated with high efficiency to
generate beams with phase singularities at optical
wavelengths~\cite{Brand1}. Thus the theoretical results obtained
above indicate the possibility of realizing amplitude modulation of
spatial light in SQDs medium.
 \vskip 2pc
\textbf{IV CONCLUSION}
 \vskip 1pc
In conclusion, the generation of LG modes from Gaussian beam has
been demonstrated in SQDs due to CPO. Both phase modulation with low
absorption of probe field and amplitude modulation with high
efficiency can be achieved, indicating that the system can be used
as effective optically addressed SLMs. Thus the SQDs system is a
promising candidate to realize spatial light modulation in a
practical way and may be useful for applications in holographic
technology and quantum information processing.

 \vskip 2pc

\begin{acknowledgments}
This work has been supported in part by National Natural Science
Foundation of China (No.10774101 and No.10974133) and the National
Ministry of Education Program for PhD.
\end{acknowledgments}

%\newpage
%\centerline{\large{\bf References}}

%\v{Z}

%\clearpage
%\begin{figure}
%\includegraphics{fig1.eps}
%\caption{}
%\end{figure}

\end{document}